# Digit classification using photonic reservoir computing based on a silicon microring resonator

Andrey A. Nikitin[1], A. A. Ershov[1], A.V. Kondrashov[1], Alexander S. Smirnov[2], Sergey S. Kosolobov[2], Anastasiya K. Zemtsova[2], Daniil S. Zemtsov[2], Alexandra I. Vergules[2], Vladimir P. Drachev[2], and Alexey B. Ustinov[1]

[1]*Department of Physical Electronics and Technology, St. Petersburg Electrotechnical University, St. Petersburg 197376, Russia*

[2]*Center for Engineering Physics, Skolkovo Institute of Science and Technology, Moscow 121205, Russia*

We demonstrate first experimental investigation on the performance of a single-node reservoir computer based on a silicon microring resonator (MRR) operating on the digit recognition task. The input layer of the reservoir is composed of a single laser, a Mach-Zehnder electro-optic modulator, which encodes intensity of the light applied to the MRR input. The input signal is transformed into a virtual high-dimensional space through thermal nonlinearity in the MRR. The MRR response is recorded with readout network consisting of a photodetector and an oscilloscope. To study the principle of operation we measure nonlinear frequency response as well as dynamic response of the MRR. The resonator demonstrates a negative shift of the resonance frequency with an increase in input power due to the dominating thermo-optic effect. In addition to the frequency shift, the MRR transmission coefficient grows at the red side of the resonance. This effect underlies the nonlinear transient dynamics at the MRR output and provides an intrinsic fading memory that are the basis for the implementation of a reservoir computer. Here a silicon MRR serves as a single nonlinear physical node. We give proof of concept demonstrations of the developed reservoir architecture by solving the classification task. The performance characteristics is evaluated with the short-term memory and the parity-check tests. Obtained results pave the way to chip-scale optical reservoirs computing.

In the past decade, a hardware implementation of reservoir computers got a lot of attention and opened new avenues for neural networks [1, 2]. In general, reservoir computing (RC) system consists of three layers. An input layer receives and prepares information. The first benefit of the physical RC is that the information can be injected in an analog form. This allows to defer the analog-to-digital conversion to a final stage [3]. A reservoir (middle) layer, typically defined by some nonlinear recurrent network dynamics, converts the input signal to a higher-dimensional state space to induce linear separability of processed data. A third layer extracts the results and produces the final output. Another important feature of the physical RC is that the training process relates to the output layer only [4].

In consistent with the classifications given in Ref. [5], physical RCs can be divided into three types by their topological structure. These are (1) the discrete physical nodes reservoirs composed of a network of memristors [6,7], spintronic nano-oscillators [8,9], optical ring resonators [10,11]; (2) the continuous-medium reservoirs utilizing the wave phenomena in a continuous medium [12]; and (3) the single-node reservoirs with delayed feedback, which provides a transformation of input signal into a virtual high-dimensional space through time division multiplexing [13]. The last type of reservoirs is especially suited for the hardware implementations, that is confirmed in many proof-of-concept experiments such as magnonic RC [14-19], bulk optoelectronic RC [20-24], and integrated photonic RC [25-30], to name a few.

Among various platforms, photonic nonlinear dynamical systems demonstrate a large potential for RC scalability and resolve conflicting demands between growing in the number of the nodes and their interconnections and decreasing in a size and an energy consumption. An elegant realization of RC node with a single silicon microring resonator (MRR) has been proposed in Ref. [27]. That work reports first experimental investigation of a silicon MRR operating as a single node RC, which successfully completes the recognition task. This RC uses two lasers: a pump laser, which encodes the data injected into the MRR, and a probe laser tuned at a different resonance wavelength and read out the MRR response. It was shown in Ref. [27] that single silicon MRR fulfils two basic requirements of suitability as a single-node RC. First, it provides the nonlinear mapping of the input data to a higher-dimensional state space. The origin of nonlinear behavior of the silicon MRR is competing dispersion effects induced by carrier and thermal nonlinearities caused by generating of the free carries through two-photon absorption [31-36]. Second, it demonstrates an intrinsic fading (short-term) memory in the system caused by the free carrier recombination and the temperature cooling.

In this work, we realize digit classification with a physical effect of nonlinear transient process, which takes place after abrupt change in the power of operating optical wave. We experimentally investigate nonlinear response of a silicon microring resonator to input optical pulsed signal from a single laser. Complex transient dynamics at the resonator output induced by the input pulses manifests itself when the laser is red-tuned from the resonance. In this operating regime, the resonator response is determined by its own output in the past that provides a transformation of the input pulsed signal into a virtual high-dimensional space. Finally, we give proof of concept of the single-node RC based on a silicon micro-ring resonator operating with a single laser, which successfully completes the digit recognition task.

The experimental resonator structure consists of a silicon microring in add-drop configuration symmetrically coupled to straight input and output waveguides (see Fig. 1). It is fabricated on a silicon-on-insulator substrate with a 2 μm buried oxide layer by electron-beam lithography and inductive-coupled plasma etching technology. Sol gel derived spin coated silica film based on tetraethylorthosilicate (TEOS) is used as an upper cladding. The resulting thickness of the cladding layer is 300 nm. The silicon waveguides forming the structure have widths of 450 nm and heights of 215 nm. The diameter of the ring is 16 μm. The distance between the ring and the straight input and output waveguides is 167 nm. The light is coupled through an out-of-plane lensed fiber to the Bragg grating coupler. The couplers have a width of 14.7 μm, a period of 655 nm, a filling factor of 0.64, and an etch depth of 70 nm. Linear tapers with length of 500 μm are used at both ends of the waveguides to match the mode diameters. Lensed fibers with anti-reflective coating used for coupling have a focal spot size of 5 μm and a working distance of 26 μm. The angle of the fibers is chosen to ensure the minimum fiber-to-fiber insertion loss at central frequency of 191 THz, which is found to be -13 dB.

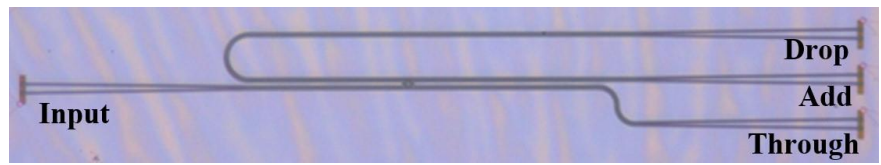

Fig. 1. Silicon microring resonator.

An experimental study of the MRR characteristics is carried out in three steps. The first step is measuring the linear resonant properties of the MRR, namely the spectrum of the resonant frequencies, the coupling coefficients, as well as the linear propagation loss. The second step involves investigating the MRR characteristics in the nonlinear operating regime. The final step is devoted to study the transient output dynamics of the MRR induced by input pulses.

As the initial step, linear transmission characteristics of the MRR are measured. High Definition Component Analyzer by Aragon Photonics is used for this purpose. The characteristics obtained from the Through and Drop ports of the MRR are shown in Fig. 2(a) and Fig. 2 (b), respectively. To ensure the linear operation, low optical input power of -10 dBm is used in the experiments.

The MRR demonstrates multiresonant response with the distance between the neighboring harmonics, i.e. the free spectral range, about 1.38 THz. The insets to Figs. 2(a) and 2(b) show the fragments of the characteristics measured in the vicinity of the resonant harmonic $f_0$ = 191.82 THz. Further, we restrict ourselves to the realization of photonic RC for this resonance harmonic.

An original characterization method is applied for an extraction the frequency-dependent attenuation and coupling coefficients from the experimental transmission characteristics [37]. The extracted dependences are shown in Figs. 2(c) and 2(d). As it is seen, the losses (α) and power coupling coefficient (κ) exhibit almost linear dependences on frequency. Its values at $f_0$ are α = 1.9 dB/mm and κ = 0.035 that correspond to loaded Q-factor of 8700. Note here that some resonances are excluded from the extraction procedure due to resonant doublet [38, 39], which makes classical estimation of Q-factor prohibitive [40].

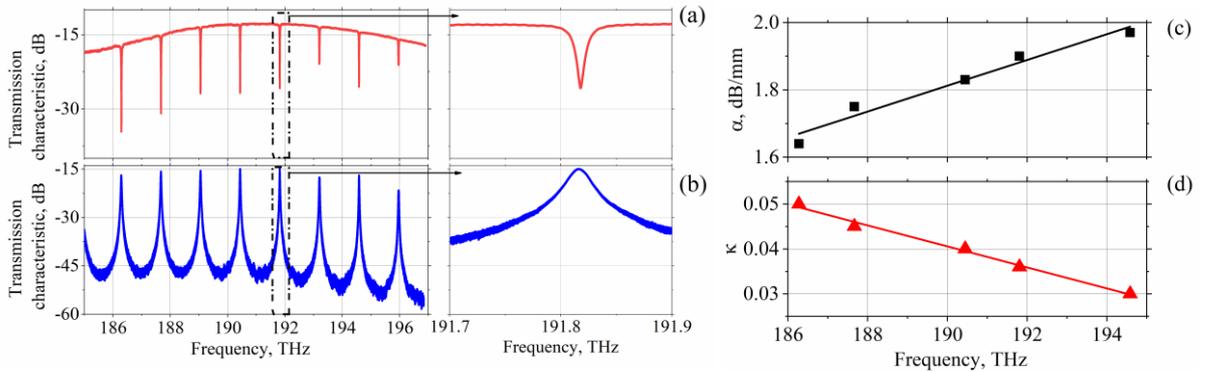

Fig. 2. Transmission characteristics observed from the Through (a) and Drop (b) ports of the MRR. The insets show the fragments of the transmission characteristics in the vicinity of the resonant harmonic of $f_0$ = 191.82 THz. The frequency dependences of the attenuation (c) and power coupling (d) coefficients.

The nonlinear transmission characteristics are measured at the next step. The experimental setup is sketched in Fig. 3(a). A 10-kHz linewidth tunable laser module Pure Photonics PPCL300 with the output power of 17 dBm is used as a source of a continuous-wave optical radiation. The laser is operated in a continuous frequency sweep regime. Frequency of the optical radiation is swept both up and down with the rate 1 GHz/s. A two-port arbitrary waveform generator (AWG) and Mach-Zehnder electro-optic modulator (EOM) serve for the control of the power applied to the MRR. A polarization controller (PC) adjusts the light polarization to enhance both the coupling efficiency and output power. The lensed fiber coupled with the Through port of the MRR is connected to output branch, which consists of two channels (see dashed lines in Fig. 3(a)). The first channel (number 1 in Fig. 3(a)) is connected to an optical spectrum analyzer (OSA), which records the MRR power-frequency response as a maximum power of the transmitted optical harmonic having a slow variation in frequency. This harmonic is generated initially with the single-frequency tunable laser described earlier. The second channel (number 2 in Fig. 3(a)) is connected to a photodetector (PD) and oscilloscope (OSC) to measure the temporal response of the MRR.

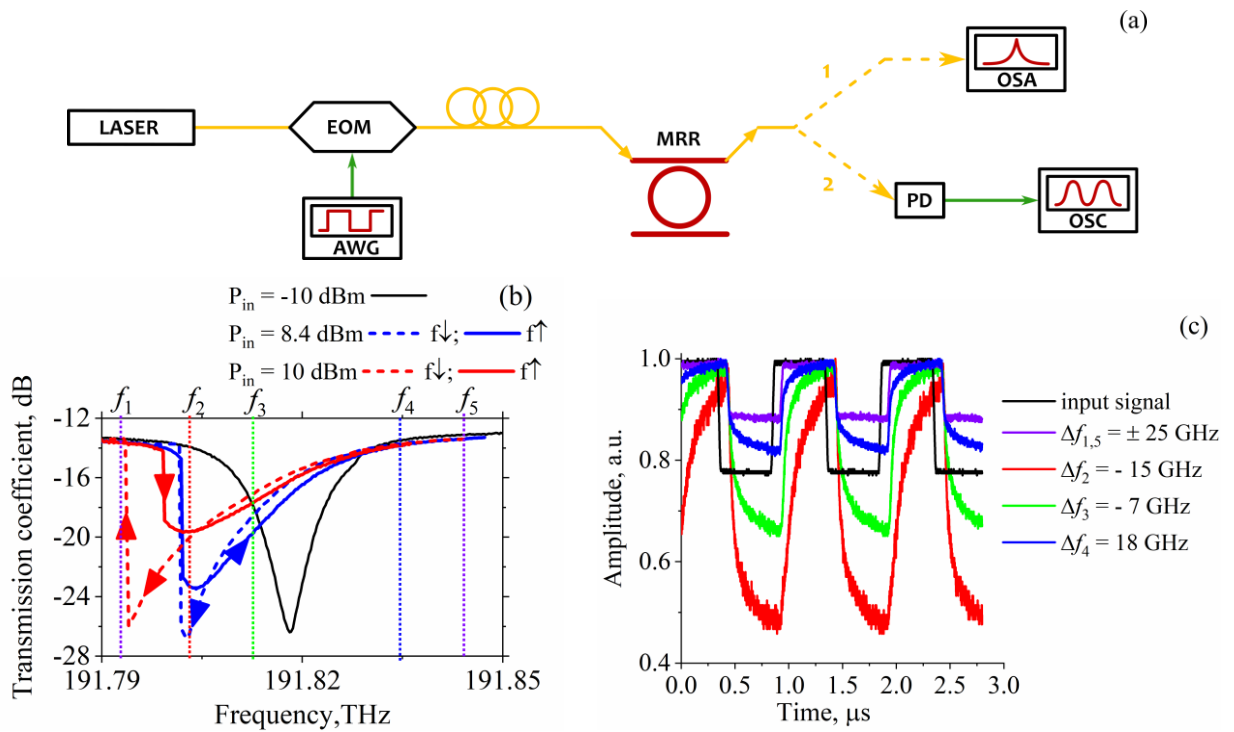

Fig. 3. Sketch of the experimental setup (a), nonlinear transmission characteristic of the MRR (b), waveforms of the input signal and temporal responses of the MRR from the Through port (c).

Then we continued with the measurements of the nonlinear transmission characteristics. The voltage $U_{EOM}$ applied to the electrical input of the EOM is used to control the optical power applied to the MRR input $P_{in}$. Namely, $U_{EOM}$ = 500 mV gives EOM transmission coefficient of $T_{EOM}$ = -27 dB, which yields $P_{in}$ = -10 dBm. Such input power ensures the linear operation of the MRR and gives the transmission characteristic shown by black solid line in Fig. 3(b). As it is seen, the obtained characteristic is identical with the one shown in the inset to Fig. 2(a). For further experiments following two values of the EOM voltage are used $U_{EOM}$ = 1500 mV and $U_{EOM}$ = 1900 mV that give $T_{EOM}$ = - 8.6 dB and $T_{EOM}$ = -7 dB and yield $P_{in}$ = 8.4 dBm and $P_{in}$ = 10 dBm, respectively. The transmission characteristic measured for $P_{in}$ = 8.4 dBm and $P_{in}$ = 10 dBm are shown in Fig. 3(b) by blue and red lines, respectively. Dashed lines in the figure show the results recorded when the laser frequency is swept down, and the solid lines show the results when the laser frequency is swept up. As it is seen, an increase in the input power leads to frequency pulling. The resonant frequency shifts downwards by 15 GHz for $P_{in}$ = 8.4 dBm and by 24 GHz for $P_{in}$ = 10 dBm. This phenomenon is associated with an increase in the refractive index caused by the dominating thermo-optic effect. In addition to the frequency pulling one sees that a growth in $P_{in}$ results in widening of the region of the hysteresis loop that corresponds to the frequency range of the bistable behavior. Namely, the increase of the input power from $P_{in}$ = 8.4 dBm up to $P_{in}$ = 10 dBm broadens the width of the hysteresis loop from 0.2 GHz to 5.7 GHz (see Fig. 3(b)).

At the third step, we studied the transient processes induced at the MRR output by the input pulses produced by the EOM. The experiments is carried out for various frequency detuning $\Delta f_i = f_i - f_0$, where $f_i$ and $f_0$ =191.82 THz are operating and resonance frequencies, respectively. As a help for eyes, positions of operating frequencies $f_i$ are shown in Fig. 3(b) by vertical lines the color of which is defined in Fig. 3(c) and will be discussed below. For the measurements, we utilizes the second channel of the experimental setup (see fig 3(a)). The pulsed signal from the AWG is applied to the EOM input. The signal has a low level of 1500 mV, a high level of 1900 mV, at the frequency of 1 MHz and duty cycle of 50 %. Note, that these parameters maintain domination of the thermal nonlinearity [32, 35]. The input signal recorded by the oscilloscope is shown in Fig. 3(c) by black line.

The first value of the frequency detuning $\Delta f_1$ = -25 GHz is selected to be beyond the range of the nonlinear response of the MRR. This detuning value allows to avoid an influence of the nonlinearity. Violet line in Fig. 3(c) shows the output signal detected by the photodetector and recorded by the oscilloscope for this frequency detuning. As it is seen, the output and input signals are linearly connected. The time shift of 80 ns between the output and input signals occurs from the signal delay in the optical fibers used in experiments. The output pulses with relatively high height and well-pronounced transient dynamics appear for $\Delta f_2$ = -15 GHz (see red line in Fig. 3(c)). Note, that the height of the output pulses is determined by the difference in the MRR transmission characteristics obtained for high ($P_{in}$ = 10 dBm) and low ($P_{in}$ = 8.4 dBm) input powers (see Fig. 3 (b). Since this difference reaches the maximum for frequency $f_2$, we observe the pulses with the highest height. Further blue detuning ($f>f_2$) provides a convergence of the transmission characteristics, thus produces a reduction in both the pulse height and the transient time with the frequency. See green line for $\Delta f_3$ = -7 GHz and blue line for $\Delta f_4$ = 18 GHz in Fig. 3(c). Finally, when the laser detuning becomes far away from the resonance frequency $\Delta f_5$ = 25 GHz, the output pulses becomes identical to the pulses observed for $\Delta f_1$ (see violet line in Fig. 3(c)). An additional test measurement shows that observed dynamics of the MRR is maintained as long as the duty cycle is greater or equal to 10 % (the pulse length greater or equal 100 ns).

The described above the nonlinear behavior opens up an avenue to exploit a single silicon MRR as a physical reservoir, in which virtual nodes are induced by a single pulsed input optical signal using time multiplexing. This idea has been shown on pioneering proof-of-concept report [30]. However, an experimental examination of such a system operating as RC is lacking. The architecture of the RC based on the single silicon MRR is the same as used for investigation of transient processes earlier.

Consider implementation of digit classification with the MRR. The first task we addressed is recognizing digits (0-9) from images. The images used for this purpose and preprocessing procedure are like those referred in works [6, 41]. The images are a 4×5-pixel black-and-white bitmaps. As an example, Figs. 4(a) and 4(b) illustrate the bitmaps for the digits "1" and "2", where black and white pixels encode as logical '0' and '1', respectively. As is shown in the insets to these figures, the images are divided into 5 rows, each row containing 4 consecutive pixels. The rows are concatenated into a 20-bit string. So that the bit sequence '01000100010001000100' encodes the bitmap for the digits "1", and bit sequence '10011110110110110000' encodes the bitmap for the digits "2". The final sequences that are fed to the RC consist of 600 digits. In these sequences each digit (0-9) appears 60 times in randomized order and is encoded with the rule discussed above. Thus, the length of the final sequences is 12000 bits.

Note that the classification problem considered is relatively simple, since the images of the same digits do not differ from each other. However, it is practically important to evaluate the recognition accuracy for noisy images. For the preparation of these images, logical negation is performed to a single random bit or to each of two random bits of the bitmaps. Using this procedure additional two binary sequences for the digits with one noisy bit as well as with two noisy bits are prepared.

For the realization of temporal multiplexing AWG generates a temporal waveform in accordance with the preprocessed binary sequence. The waveform levels are chosen to be the same as in previous experiments. Namely, the low level of 1500 mV encodes as logical '0' and high level of 1900 mV encodes as logical '1'. The bit length of 100 ns (the bit-rate of 10 Mbit/s) ensures domination of the thermal nonlinearity. The constructed waveform is applied to the electrical input of the EOM, which encodes light intensity of the pump laser. As an example, Fig. 4(c) shows the region of the waveform corresponding to the digit "2". The light pulses injected into the MRR induces the transient dynamics that represents the reservoir output. The MRR response is recorded with the readout network consisted with a photodetector and an oscilloscope.

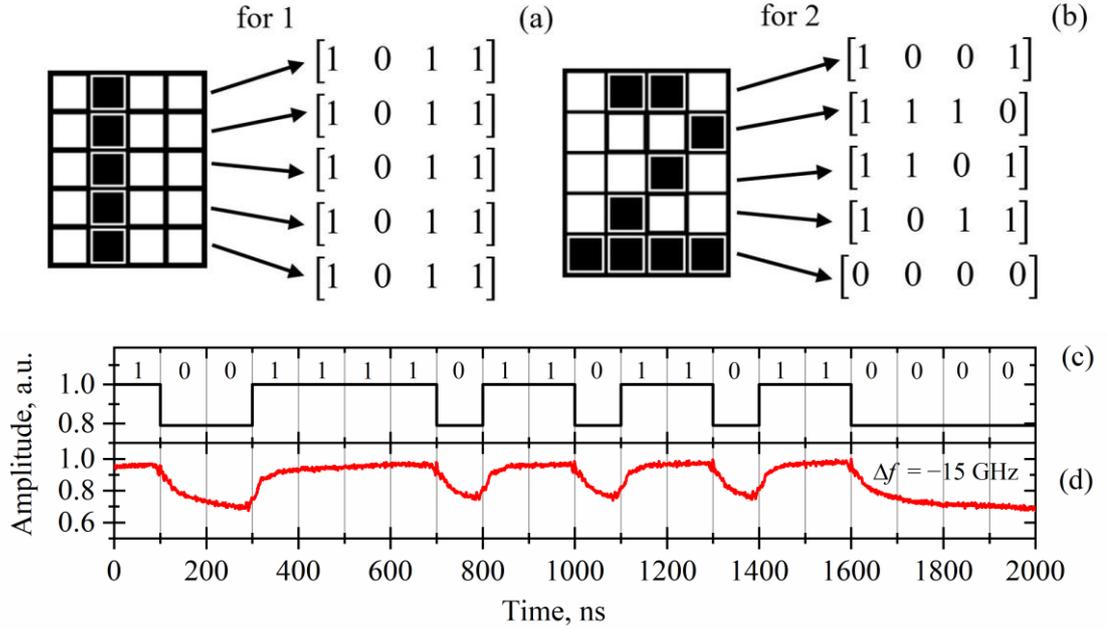

Fig. 4. Bitmaps for the digits "1" (a) and "2" (b); the waveforms of input (c) and output (d) signals corresponding to the digit "2".

As is discussed above, the response of the MRR is specified by the frequency detuning value. Thus, the experiments are carried out for the various frequency detuning. Fig. 4(d) shows the output waveform received by the oscilloscope for detuning of $\Delta f = -15$ GHz in response to the input signal presented in Fig. 4(c). Below, we restrict ourselves to investigation of the RC operation for this frequency detuning. As it is seen in Fig. 4(d), the MRR response satisfies the key properties of the reservoir. First, the recurrent nonlinear dynamics of the MRR provides transformation of the input signal to a high-dimensional state space. Second, it exhibits a short-term memory provided by the free carrier recombination and the temperature cooling. Note, the outputs are influenced by the inputs from the recent pass, but insensitive to the information from the far past.

The readout network is trained using linear regression as discussed below. To increase the dimensionality of the reservoir, the MRR output is sampled into $N$ 'virtual' nodes for each image with separation $2000/N$ ns, where 2000 ns is the running time of each image. The voltage values at these time intervals correspond to the RC states. Obtained vector of the states is split into two equal data sets. Each data set consists of 300 images. The first data set is used for the readout function training. After training, the second data set is used for the classification and testing the recognition accuracy. The dependences of the recognition accuracy on the number of virtual nodes $N$ obtained for the noiseless images as well as for images with one and two noisy bits are shown in Figs. 5(a), 5(b) and 5(c), respectively. As is seen from these figures, $N = 30$ virtual nodes provide accuracy of 92 % even for the noisy images.

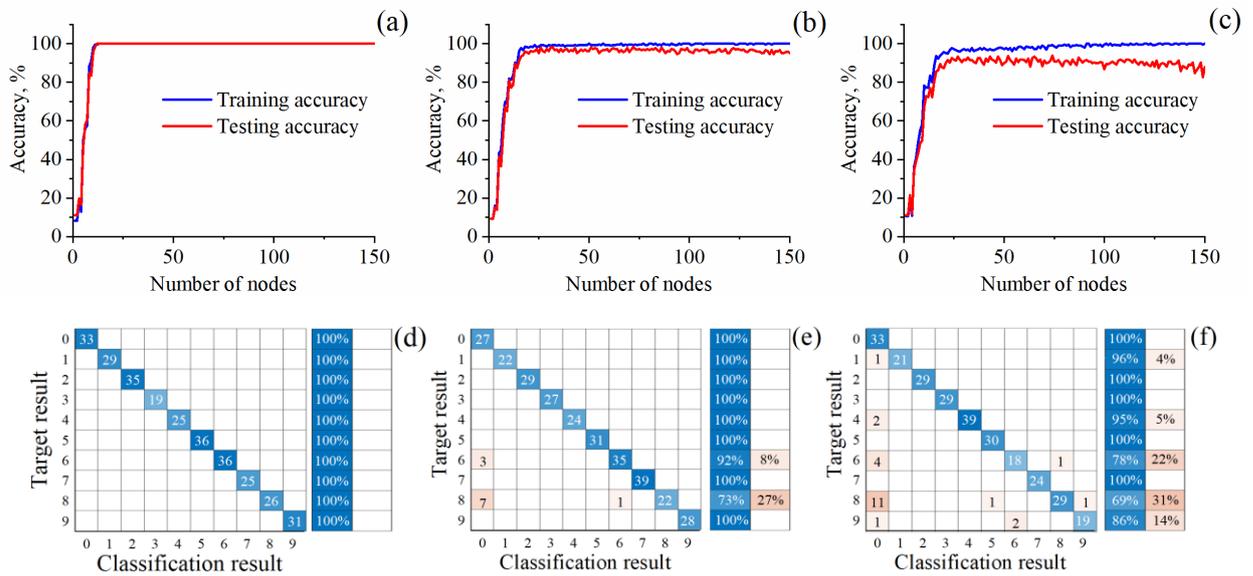

Fig. 5. The recognition accuracy of the testing (blue lines) and training (red lines) data sets on the number of the virtual nodes $N$ for the noiseless images (a) as well as for images with one (b) and two (c) noisy bits. False color confusion matrices for the noiseless images (d) as well as for images with one (e) and two (f) noisy bits.

Figs. 5(e-g) visualize false color confusion matrices obtained for $N = 30$. These figures quantitatively characterize the efficiency of recognizing noiseless and noisy images. The first axis in figures shows the obtained classification results, and the second axis indicates the target outputs. The diagonal elements of the matrices show the number of correctly identified images in the sequence. The off-diagonal elements of the matrices show the number of images referred to another class.

Next we evaluate two conventional metrics that are the short-term memory (STM) and parity-check (PC) tests. The STM test is a memory recall task, where the targets are defined as the input values at some delay time. The PC test is a non-linearly separable task, which requires both fading memory and nonlinearity. These tests are consistent throughout the literature [12, 42-45]. Therefore, we do not discuss them here. The method of evaluating STM and PC capacities ($C_{STM}$ and $C_{PC}$) consists of training and subsequent testing of RC. Since both benchmark tests operate with binary sequences, only one input sequence is needed to compute both $C_{STM}$ and $C_{PC}$ capacities. In our experiments, a 2000-value binary series is fed into the MRR. The MRR output is sampled into $N = 30$ virtual nodes for each bit time interval with separation $100/N$ ns, creating a reservoir state vector, where 100 ns is the pulse length. The data set of 1000 outputs, each of which involving 30 virtual nodes, are used for the training process. This number is sufficient to saturate the values of the capacities as well as to ensure that the regression models are not overfit. The remaining 1000 outputs are used for the capacity's evaluation.

The success of the linear regression models to reconstruct the desired targets is characterized by the correlation coefficient $r$. This coefficient is a measure of the reproducibility of target reconstruction for each input data injected before the current pulse. The results of the STM and PC tests for the reservoir under investigation are shown in Fig. 6. As it is seen from the figures $C_{STM} = 2.39$ and $C_{PC} = 0.91$. These capacities are comparable to the results obtained for other reservoir architectures based a single spin-torque oscillator [8] as well as a magnonic active ring oscillator [46, 47].

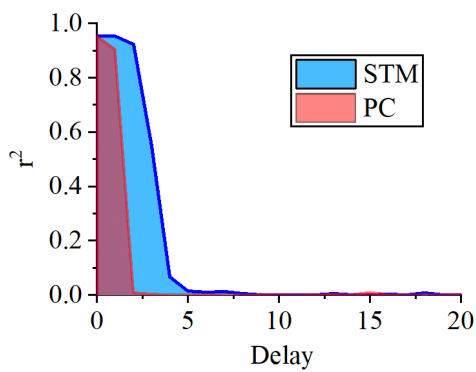

Fig. 6. Correlation coefficients versus delay for the STM and PC tasks. The areas of blue and red colors under the correlation coefficients represent the capacities for the benchmark tests.

In conclusion, the obtained results demonstrate proof-of-concept of the reservoir computing based on a single silicon microring resonator and a single pump laser. To demonstrate the principle of the reservoir operation the linear and nonlinear transmission characteristics of the resonator are investigated. It is shown that resonator demonstrates linear response for the operating power below -10 dBm. The frequency dependences for the attenuation and power coupling coefficients are extracted from linear transmission characteristics using an original characterization method. An increase in the input power leads to negative frequency shift associated with the dominating thermo-optic effect. The input power of 2 dBm yields the nonlinear frequency shift of 2.4 GHz. It is shown that an abrupt change in the input power of operating optical wave by 1.6 dBm induces the nonlinear transient process at the resonator output, which strongly depends on the frequency detuning. These results are promising for hybrid fabrication of different neuromorphic devices as photonic integrated circuits including lasers, microring resonators, and photodetectors.

The research in St. Petersburg Electrotechnical University was funded by the Ministry of Science and Higher Education of the Russian Federation under Grant No. FSEE-2025-0008.

## AUTHOR DECLARATIONS

**Conflict of Interest**

The authors have no conflicts to disclose

**Data availability**

The data that support the findings of this study are available from the corresponding author upon reasonable request.